\documentstyle[twoside,fleqn,espcrc2,epsfig]{article}

\newcommand{\AmS}{{\protect\the\textfont2
  A\kern-.1667em\lower.5ex\hbox{M}\kern-.125emS}}

\centerline {\bf ~~~~~~~~~~~~~~~~~~~~~~~~~~~~Hints of
$\mu$+N$\rightarrow$$\mu$+N+$\mu^{+}$$\mu^{-}$ process observation
with the MACRO Experiment} 
\centerline {\bf ~~~~~~~~~~~~~~~~~~~~~~~G. Battiston$i^{*}$ and E. Scapparon$e^{+}$  
for the MACRO Collaboration}
\centerline{~~~~~~~~~~~~$~^{*}$INFN Milano, via Celoria  16,
~20133,~Italy~~~~~~~~~~~~~~~~~~~~~~~~~~~~~~~~~~~~~~~~~~~~~~~~~~~~} 
\centerline{~~~~~~~~~~~~~~~~~~~~~~~~$~^{+}$ INFN - Laboratori Nazionali del Gran Sasso, S.S.17 km 18+910, 61070, Assergi (AQ), Italy}      
\centerline{ \bf ~~~~~~~~~~Talk at Xth International Symposium on Very High}
\centerline{ \bf ~~~~~~~~~~Energy Cosmic Ray Interaction}
\centerline{\bf ~~~~~~~~~~July 12 - 17, 1998}
\centerline{\bf ~~~~~~~~~~To be published in Nucl. Phys. B, Proc. Suppl.}

\title{Hints of 
$\mu$+N$\rightarrow$$\mu$+N+$\mu^{+}$$\mu^{-}$ process observation
with the MACRO Experiment}                                        

\author{G.Battistoni\address{INFN Milano, via Celoria    ,Italy} 
and E. Scapparone\address{INFN, LNGS
        S.S.17 km 18+910, 61070, Assergi (AQ), Italy}
       for the MACRO Collaboration
\thanks{For the complete list of
the Collaboration see the paper "Relevance of the hadronic interaction
model in the interpretation of multiple muon data as detected with MACRO
experiment" by O. Palamara at these proceedings}
       }
\begin{document}

\begin{abstract}
We present the analysis of the distribution of the distance of muon pairs
detected deep underground by the MACRO experiment. A detailed analysis
of the low separation region pointed out an excess of experimental data
with respect to the Monte Carlo expectation. The effect is discussed taking 
into account the 
$\mu$+N$\rightarrow$$\mu$+N+$\mu^{+}$$\mu^{-}$ process.
\end{abstract}\vspace{0.2cm}

\maketitle

\section{Introduction}
The distribution of the muon pair separation deep underground 
(decoherence function) 
plays an important role in the study of Cosmic Ray
physics in underground experiments, being sensitive to some
aspect of the hadronic interaction model. Taking advantage of the 
large area, MACRO\cite{macro} can
access the largest separation, exploring thus the highest $P_{t}$ regions.
Moreover,
the capability of the MACRO detector to resolve very close tracks
permits the extension of the decoherence analysis to a distance 
region hardly studied in the past.


\section {Analysis of the behaviour at small distances}

We have analyzed about $3.4 \cdot 10^{5}$ events, corresponding 
to 7732 hr live time using the lower part of the apparatus. 
These events were submitted to the following selection criteria:\\
- Zenith angle smaller than $60^0$: this choice is dictated 
by our limited knowledge of the Gran Sasso topographical map for 
high zenith angles. Furthermore, we cannot disregard the atmosphere
curvature for larger zenith angles, while at present our
simulation model does not include that.\\
- The number of streamer tube hits out of
track smaller than 45. This selection aims to eliminate possible 
misleading track reconstruction in events produced by noise in the
streamer tube system and/or e.m. interactions inside or in the neighbourhood 
of the apparatus.\\
- Track pairs which survive a loose parallelism cut:
this rejects hadrons from photonuclear interactions close to
the detector. This selection also eliminates tracks reconstructed 
using hits due to e.m. interactions in events survived to the previous cut.

The last cut is not completely efficient in rejecting muon
tracks originating from local particle production because 
the angle between these tracks may fall within the limits 
imposed by the parallelism cut.
Moreover, these limits cannot be further reduced since the average
angular divergence due to multiple muon scattering in the rock 
overburden, is about 1$^\circ$ at the MACRO depth.

A further selection was applied in order to reduce these effects. 
We computed, for each muon track in the wire view, the ratio 
R between the number of streamer tube planes fired by the muon to the 
number of expected fired planes considering the track direction. 
Only tracks with $R \geq 0.75$ were accepted.
The application of this cut (hereafter the C4 cut) in the wire view 
alone is a good compromise between the rejection capability of the algorithm 
and the loss of events due to the unavoidable inefficiency of the streamer 
tube system. We found that in the wire projective view the probability 
to reject a muon track for a casual alignment contiguous 
inefficient planes is 2.0\%.

To test the ability of this cut to reject hadronic tracks,
we used FLUKA \cite{fluka} to simulate 3028 hr of live time 
in which muons were accompanied by hadronic products of photonuclear
interactions in the 10 m of rock surrounding the detector.
We found that the parallelism cut alone provides a rejection efficiency
of about 54.6\% of the pair sample, while the addition of cut C4 enhances the 
rejection to 95.9\%. Furthermore the effect of hadron contamination, 
is very small, contributing less than 1\% in the overall muon pair sample.
After the overall application of these cuts, the number of 
unambiguously associated muon pair tracks survived is 355,795.
\begin{figure}[tb]
\vskip -2.2 cm
\begin{center}
\mbox{\epsfysize=75mm 
      \epsffile{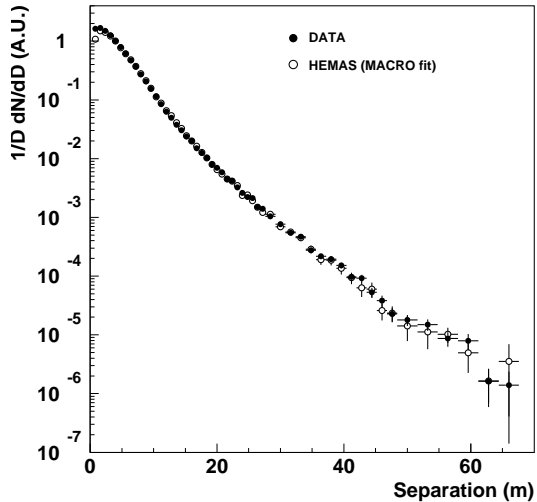}}
\end{center}
\vskip -2.2 cm
\caption{\em Comparison between the experimental 
decoherence function with the HEMAS expectation. \label{f:folded}}
\vskip -0.8cm
\end{figure}      
Fig. 1 shows the comparison between real data and HEMAS expectation:
a general good agreement is found, showing the HEMAS capability
in reproducing the muon bundle features deep underground.
Nevertheless a relevant discrepancy at low distances is observed,
indicating a possible physical source for this excess in the real data.
For istance, considering separation smaller than 80 cm, an excess of 55$\%$ 
is observed in the real data(Table 1).
The process of muon pair production by muons in the rock 
is a natural candidate. As pointed out in \cite{olga}, at the
typical muon energy involved in underground analyses 
($E_{\mu} \sim$ 1 TeV), if 
very large energy transfer are considered,
the cross section of such a process is non negligible with respect to the
e$^+$e$^-$ pair production. The key point is that at 
high v=$E_{\gamma}$/$E_{\mu}$, the most important Feynmann diagram
is rather insensitive to the ratio ($m_{e}$/$m_{\mu}$$)^{2}$.
An analytic expression for the muon pair production cross section is
given in \cite{olga,kelner}. In order to test the hypothesis, such
cross section 
has been included in the muon transport code PROPMU\cite{lipari_stanev}.
Assuming a muon flux with energy spectrum $E^{-3.7}$ 
with a minimum muon energy $E_{\mu}^{min}$ = 1.2 TeV at the surface
and considering the actual mountain profile, we generated a sample of 
$10^{7}$ muons corresponding to 3666 h of live time.
About $\sim 3.0 \cdot 10^{6}$ muons survived to the MACRO level,
5360 of which derived from muon pair production processes.
The average distance of these muon pairs is $(128\pm1)$ cm and
their average energies are $(657\pm14)$ GeV and $(145\pm3)$ GeV for,
the main muon and the secondary muon samples respectively.
We propagated the muons surviving to the MACRO level 
through GEANT simulation and we applied to these events the same cuts 
specified in Section 2. Finally, the number of events was normalized
to the live time of real data.

In Table \ref{t:tab1} we report the number of weighted muon pairs
in the first bins of the experimental and simulated decoherence 
distributions expressed as percentage with 
respect to the population in the bin of the distribution maximum.
The effect of standard cuts, of the C4 cut and of the subtraction 
of the muon pair production process are shown in order. 
In each case, we indicate in percentage the bin populations with 
respect to the peak of the distribution and the discrepancy with
respect to the Monte 
Carlo predictions.

\begin{figure}[tb]
\vskip -2.2 cm
\begin{center}
\mbox{\epsfysize=75mm 
      \epsffile{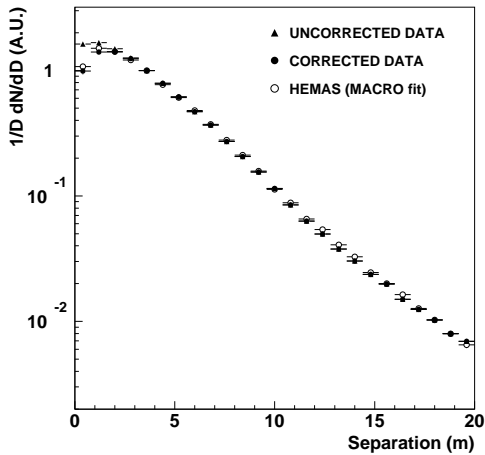}}
\end{center}
\vskip -2.2 cm
\caption{\em The low distance region of the experimental 
decoherence function before and after the subtraction of the secondary 
muon sample and comparison with the Monte Carlo simulation. \label{f:deco_clean}}
\end{figure}

In Fig. \ref{f:deco_clean} we compare the simulated decoherence curve
with the experimental data corrected for the muon pair production effect.
Despite the approximation introduced in our test, 
it seems that the proposed muon pair production process
can account for the most of observed discrepancy in the low distance range.

\begin{table*}[t]
\setlength{\tabcolsep}{1.5pc}
\newlength{\digitwidth} \settowidth{\digitwidth}{\rm 0}
\catcode`?=\active \def?{\kern\digitwidth}
\caption{\em Number of weighted muon pairs in the first bins of
the experimental and simulated decoherence distributions.
The discrepancy is the percentage difference between experimental
and Monte Carlo values normalized to the distribution maximum 
(last column). \label{t:tab1}}
\begin{tabular*}{\textwidth}{@{}l@{\extracolsep{\fill}}ccccc}
\hline
& 0--80 cm & 80--160 cm & 160--240 cm & 240--320 cm & 320--400 cm \\ 
& & & & & (max) \\
\hline \hline
Exp. Data & 5528 & 12491 & 17569 & 20514 & 20816 \\
\hline
MC Data   & 5154 & 21417 & 33573 & 40367 & 42679 \\
\hline
Discrepancy & (55$\pm$2)\% & (16$\pm$2)\% & (6$\pm$1)\% & (4$\pm$1)\% & \\ 
\hline \hline
Exp. Data + C4 & 3612 & 11128 & 16535 & 19597 & 19977 \\
\hline
MC Data + C4& 4848 & 20346 & 31932 & 38425 & 40660 \\
\hline
Discrepancy & (34$\pm$2)\% & (10$\pm$2)\% & (6$\pm$2)\% & (4$\pm$2)\% & \\ 
\hline \hline
Exp. Data + C4 & 2193 & 9264 & 15462 & 19190 & 19842 \\
$\mu$ pair subtraction & & & & & \\
\hline
Discrepancy & (8$\pm$7)\% & (7$\pm$3)\% & (0$\pm$2)\% &  (2$\pm$2)\% & \\ 
\hline
\end{tabular*} 
\end{table*}
\section {Discussion and conclusion}

We measured the underground decoherence function
using high energy muons ($E_{\mu}> 1.3~TeV$) 
up to a maximum distance of about 70 m.

The capability to resolve very close muon tracks permitted the investigation
of the behaviour of the decoherence function for small separations.
Apart from the negligible contamination of hadro-production by muons, 
we found strong hints of a contribution coming from 
muon pair production by muons. The inclusion 
of this process in the simulation reproduces, both in a qualitative and
quantitative way,
the experimental data.

\end{document}